\documentclass[twocolumn,aps,prl,epsf,showpacs,superscriptaddress]{revtex4}
\usepackage{amsmath}
\usepackage{amsfonts}
\usepackage{epsfig}
\usepackage{epsf}
\usepackage{array}

\begin{document}

\title{Doped Mott Insulators in (111) Bilayers of Perovskite Transition-Metal Oxides with a Strong Spin-Orbit Coupling}

\author{Satoshi Okamoto}
\altaffiliation{okapon@ornl.gov}
\affiliation{Materials Science and Technology Division, Oak Ridge National Laboratory, Oak Ridge, Tennessee 37831, USA}

\begin{abstract}
The electronic properties of Mott insulators realized in (111) bilayers of perovskite transition-metal oxides are studied. 
The low-energy effective Hamiltonians for such Mott insulators are derived in the presence of a strong spin-orbit coupling. 
These models are characterized by the antiferromagnetic Heisenberg interaction and the anisotropic interaction 
whose form depends on the $d$ orbital occupancy. 
From exact diagonalization analyses on finite clusters, the ground state phase diagrams are derived, including 
a Kitaev spin liquid phase in a narrow parameter regime for $t_{2g}$ systems. 
Slave-boson mean-field analyses indicate the possibility of novel superconducting states 
induced by carrier doping into the Mott-insulating parent systems, 
suggesting the present model systems as unique playgrounds for studying correlation-induced novel phenomena. 
Possible experimental realizations are also discussed. 
\end{abstract}

\pacs{71.27.+a, 74.20.-z}
\maketitle



Competition and cooperation between Mott physics and the relativistic spin-orbit coupling (SOC) 
have become a central issue in condensed matter physics. 
As these two effects become comparable, 
$4d$ and $5d$ transition-metal oxides (TMOs) could be ideal platforms to explore novel phenomena originating from 
such interactions. 
This brought considerable attention to iridium oxides \cite{Kim08,Pesin10,Wang11}. 
%
%
Of particular interest is $A_2$IrO$_3$ ($A$=Li or Na) where Ir ions form the honeycomb lattice. 
Density-functional-theory calculations for Na$_2$IrO$_3$
predicted the quantum spin Hall effect \cite{Shitade09}. 
Alternatively, with strong correlation effects, 
the low-energy properties of $A_2$IrO$_3$ could be described by a combination of 
pseudodipolar interaction and Heisenberg interaction \cite{Chaloupka10}, 
called Kitaev-Heisenberg model \cite{Kitaev06}, 
which is a candidate for realizing $Z_2$ quantum spin liquid (SL) states. 
However, later experimental measurements confirmed a magnetic long-range order \cite{Singh10,Liu11,Ye12} in Na$_2$IrO$_3$ 
possibly because of longer-range magnetic couplings \cite{Kimchi11,Singh12,Choi12}. 
%
The effect of carrier doping into the Kitaev-Heisenberg model was also studied \cite{You11,Hyart12}.

\begin{figure}[tbp]
\includegraphics[width=0.7\columnwidth]{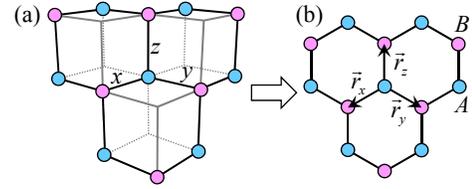}
\caption{(Color online) Buckled honeycomb lattice realized in a (111) bilayer of the cubic lattice. 
$x,y$ and $z$ in (a) indicate the cubic axes and the spin components in the Kitaev interaction 
on the buckled honeycomb lattice shown in (b).}
\label{fig:structure}
\end{figure}

Interacting electron models on a honeycomb lattice have long been theoretical targets 
for realizing novel phenomena such as 
the quantum Hall effect without Landau levels \cite{Haldane88} and the spin Hall effect with the SOC \cite{Kane05}. 
The spin Hall effect could also be generated by correlations without the SOC \cite{Raghu08}. 
Yet, experimental demonstrations for such correlation-induced phenomena remain to be done. 
Recently, artificial bilayers of perovskite TMOs grown along the [111] crystallographic axis, 
where transition-metal ions form the buckled honeycomb lattice (Fig.~\ref{fig:structure}), 
were proposed as new platforms to explore a variety of quantum Hall effects \cite{Xiao11,Ruegg11,Yang11}. 
This proposal was motivated by the recent development in synthesizing 
artificial heterostructures of TMOs \cite{Hwang12}. 
TMO heterostructures have great tunability over fundamental physical parameters, 
including the local Coulomb repulsion, SOC, and carrier concentration.
%
%
However, the effect of correlations to possible novel phenomena 
near Mott insulating states with a strong SOC remains to be explored.

Here, we address the correlation effects in TMO (111) bilayers with a strong SOC. 
Specifically, we consider $t_{2g}^5$ systems and $e_{g}^{1,3}$ systems for which 
the low-energy electronic properties could be described in terms of $S=1/2$ isospins \cite{t2g1}. 
We derive the effective Hamiltonians for such Mott insulators and analyze them numerically and analytically. 
The effective Hamiltonian for $t_{2g}^5$ has the form of the Kitaev-Heisenberg model \cite{Chaloupka10}, 
but the SL was found to exist only in a small parameter regime. 
On the other hand, the effective Hamiltonian for $e_{g}^{1,3}$ has the Ising-type anisotropy, 
thus the SL is absent. 
The effect of carrier doping is analyzed using slave-boson mean-field (SBMF) methods 
including {\it Ans{\"a}tze} which reduce to exact solutions at limiting cases of zero doping. 
It is shown that carrier doping makes the physics of our model systems 
more interesting by inducing unconventional superconducting states, 
most likely $d+id$ paring which breaks time-reversal symmetry.

{\it Effective models}.--- 
We start from a multiband Hubbard model with $t_{2g}$ orbitals or $e_g$ orbitals. 
In both cases, only the nearest-neighbor hoppings are considered, and the hopping amplitude is 
derived from the Slater-Koster formula \cite{Slater54} with oxygen $p$ orbitals located between the neighboring two $d$ orbitals. 
The explicit forms of the multiband Hubbard models 
are given in Ref.~\cite{supplement}. 

%
The low-energy effective Hamiltonian for $t_{2g}^5$ systems is 
derived from the second-order perturbation processes with respect to the transfer terms and by 
projecting the superexchange-type interactions onto the isospin states for $J_{eff}^z = \pm 1/2$ \cite{supplement}: 
\begin{eqnarray}
| J_{eff}^z = \sigma \rangle = \frac{1}{\sqrt{3}} \{ i | a, -\sigma \rangle - \sigma | b, -\sigma \rangle + i \sigma |c, \sigma \rangle\}. 
\end{eqnarray}
Here, 
$a$, $b$ and $c$ are the $t_{2g}$ multiplet given by 
$| a \rangle = | yz \rangle$, $| b \rangle = | xz \rangle$ and $| c \rangle = | xy \rangle$, respectively. 
The effective interaction between sites $\vec r$ and $\vec r'$ along the $\gamma$ bond (see Fig.~\ref{fig:structure}) reads 
\begin{eqnarray}
H_{\vec r \vec r'}^\gamma = - J_0 + J_H \vec S_{\vec r} \cdot \vec S_{\vec r'} + J_K S_{\vec r}^\gamma S_{\vec r'}^\gamma. 
\label{eq:Ht2g}
\end{eqnarray}
$J_0=\frac{1}{27} (15 r_1 + 5 r_2 + 4 r_3)$, 
$J_H=\frac{8}{27} (3r_1 + r_2 + 2 r_3)$, 
$J_K=\frac{4}{9} (r_1 -  r_2)$, 
where 
$r_1 = t_\pi^2/(U-3I)$, $r_2 = t_\pi^2/(U-I)$, $r_3 = t_\pi^2/(U+2I)$. 
%
Here, both Heisenberg and Kitaev terms have positive sign, i.e., antiferromagnetic (AFM) \cite{Jackeli09}.

%
For $e_g$ systems in the (111) bilayers, 
the SOC is activated through the virtual electron excitations to 
the $t_{2g}$ multiplet under the trigonal $C_{3v}$ crystalline field \cite{Xiao11,supplement}. 
%
%
%
Using the basis 
$| \alpha \rangle = | 3z^2-r^2 \rangle$ and $| \beta \rangle = | x^2-y^2 \rangle$, 
a low-energy Kramers doublet for $e_g^1$ is given by 
\begin{eqnarray}
| \boldmath{\sigma} \rangle = \frac{1}{\sqrt{2}} \{ | \alpha, \sigma \rangle + i \sigma | \beta, \sigma \rangle \}, 
\label{eq:egdoublet}
\end{eqnarray}
where the spin quantization axis is taken along the [111] crystallographic axis. 
For $e_g^3$, the $+$ sign in Eq.~(\ref{eq:egdoublet}) is replaced by the $-$ sign. 
This doublet can be gauge transformed to 
$\frac{1}{\sqrt{2}} \{ |  3x^2-r^2, \sigma \rangle + i \sigma | y^2-z^2, \sigma \rangle \}$ and 
$\frac{1}{\sqrt{2}} \{ |  3y^2-r^2, \sigma \rangle + i \sigma | z^2-x^2, \sigma \rangle \}$ with trivial phase factors. 
Thus, the effective interaction is expected to be symmetric with respect to the bond direction. 
Following the same procedure for the $t_{2g}^5$ systems, 
the effective interaction between sites $\vec r$ and $\vec r'$ is derived as 
\begin{eqnarray}
H_{\vec r \vec r'} = - J_0 + J_H \vec S_{\vec r} \cdot \vec S_{\vec r'} - J_I S_{\vec r}^z S_{\vec r'}^z. 
\label{eq:Heg}
\end{eqnarray}
Here, 
$J_0=\frac{1}{8} (3 r_1 + 2 r_2 + r_3)$, 
$J_H=\frac{1}{2} (r_1 + r_3)$, 
$J_I=\frac{1}{2} (2 r_1 - r_2 - r_3)$, 
where 
$r_1 = t_\sigma^2/(U-3I)$, $r_2 = t_\sigma^2/(U-I)$, $r_3 = t_\sigma^2/(U+I)$. 
%
Now the anisotropic term is described as a ferromagnetic (FM) Ising interaction. 
This comes from the fact that the total $S^z$ is conserved in the model [see Eq.~(\ref{eq:egdoublet}) 
and Ref.~\cite{supplement}]. 

The effect of the direct $dd$ transfers \cite{Khaliullin05,Chaloupka12}, 
termed $t_{\delta}$ after $(dd\delta)$ bonding, for both the models 
is discussed in Ref.~\cite{supplement}. 

{\it Undoped cases}.---
%
%
Here we discuss the AFM Kitaev-AFM Heisenberg (AKAH) model for the $t_{2g}^5$ system and 
the FM Ising-AFM Heisenberg (FIAH) model for the $e_g^{1,3}$ system 
using the parametrization $J_H=1-\alpha$ and $J_{K,I}=2 \alpha$. 
As $J_H$ and $J_K$ have the same sign, the direct transition is expected between 
the N{\'e}el AFM at small $\alpha$ and the Kitaev SL at large $\alpha$ for the AKAH model. 
For the FIAH model, the planar N{\'e}el AFM is expected at small $\alpha$ 
and the FM with the spin moment in the [111] direction at large $\alpha$. 


\begin{figure}[tbp]
\includegraphics[width=0.8\columnwidth]{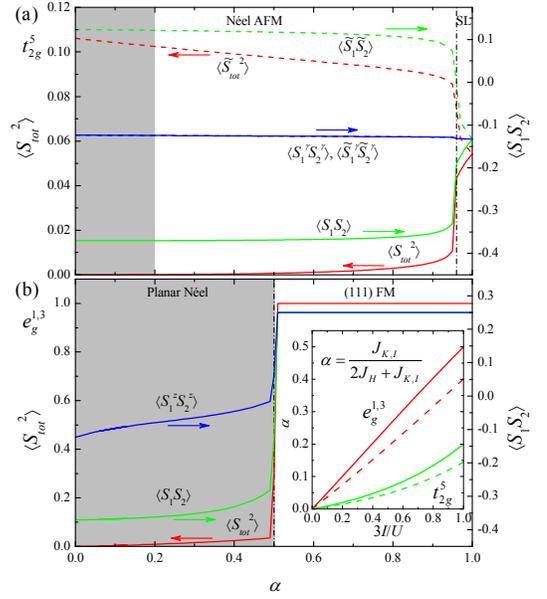}
\caption{(Color online) Lanczos exact diagonalization results, 
squared total spins (normalized to its value in the fully polarized FM state) 
and the nearest-neighbor spin correlations, for $t_{2g}^5$ model (a) and $e_g^{1,3}$ model (b) 
obtained on 24-site clusters as a function of $\alpha$. 
Solid (dashed) lines correspond to original (rotated) spin basis.
Vertical dash-dotted lines are first-order phase boundaries. 
Shaded areas are the parameter ranges for $0<3I<U$ with $t_\delta=0$. 
Inset: Controlling parameter $\alpha$ for both $t_{2g}^5$ and $e_{g}^{1,3}$ models 
as a function of $3 I/U$. 
Dashed lines include $t_\delta=0.1 t_\pi$ or $t_\delta=0.1 t_\sigma$. 
}
\label{fig:lanczos}
\end{figure}

We now employ the Lanczos exact diagonalization for the model Hamiltonians [Eqs.~(\ref{eq:Ht2g}) and (\ref{eq:Heg})] 
defined on a 24-site cluster with the periodic boundary condition. 
This cluster is compatible with the four-sublattice transformation \cite{Chaloupka10} 
which changes the original spin $S$ to $\widetilde S$. 
Numerical results shown in Figs.~\ref{fig:lanczos} (a) and \ref{fig:lanczos} (b) confirm the above considerations. 
Yet, the SL regime is found to be rather narrow for the AKAH model with the critical $\alpha_c \sim 0.96$ separating 
it from a magnetically ordered phase. 
%
%
For the FIAH model, the phase transition takes place at $\alpha = 0.5$
separating the (111) FM phase and the planar N{\'e}el AFM phase. 
In Refs.~\cite{Chaloupka12,unpublished}, the hypothetical Kitaev-Heisenberg models with different signs of interactions are studied.

Natural questions arise, such as where is the ``physical'' parameter range, i.e., $U > 3I$, 
and can $t_{2g}$ systems realize the Kitaev SL phase? 
Now, rewriting $J_{K,I}$ and $J_H$ as $J_{K,I} = 2 J \alpha$ and $J_H = J (1- \alpha)$, respectively, 
with $J$ the normalization, 
one obtains $\alpha = J_{K,I}/(2 J_H + J_{K,I})$. 
In the inset of Fig.~\ref{fig:lanczos}, we plot $\alpha$ for both the $t_{2g}^5$ and the $e_g^{1,3}$ models as a function of $I/U$. 
It is shown that $\alpha$ does not exceed $1/5$ for the AKAH model and 
$1/2$ for the FIAH model; thus both cases fall into the N{\'e}el ordered regime. 
The effect of the direct $dd$ transfers is found to merely suppress the anisotropic interactions, 
as seen as dashed lines. 
Thus, additional interactions, such as magnetic frustrations, are necessary 
to realize the Kitaev SL phase in $t_{2g}$ systems to suppress $J_H$.

{\it Slave-boson mean-field theory}.---
Although the Kitaev SL phase is outside the ``physical regime'' for Mott-insulating systems, 
there could emerge novel electronic states by carrier doping \cite{You11,Hyart12}. 
As the two models are reduced to the $tJ$ model on the honeycomb lattice at $\alpha \rightarrow 0$, 
one possible candidate is the singlet superconductivity (SC) with the broken time-reversal symmetry, 
so-called $d+id$ \cite{Black07}. 
In the opposite limit of the AKAH model, novel SC states could be stabilized in connection to the $Z_2$ SL.  
For the FIAH model, on the other hand, 
the triplet ($p$) SC states may emerge. 
Here, we examine these possibilities using a SBMF theory. 

First, we introduce a SBMF method that can be applied for Ising-like anisotropic interactions. 
An $S=1/2$ spin operator for a Kramers doublet is described by fermionic spinons $f_{\sigma}$ as 
$S_{\vec r}^\gamma = \frac{1}{2} f_{\vec r \sigma}^\dag \tau_{\sigma \sigma'}^\gamma f_{\vec r \sigma'}$, 
with the local constraint $\sum_\sigma f_{\vec r \sigma}^\dag f_{\vec r \sigma} =1$ 
and $\hat \tau^\gamma$ being a Pauli matrix. 
%
Now, a spin quadratic term can be decoupled into several different channels as 
\begin{eqnarray}
S_{\vec r}^\gamma S_{\vec r'}^\gamma \!\!\! &=& \!\!\!
-\frac{1}{8} (\Delta_{\vec r \vec r'}^* \Delta_{\vec r \vec r'} + \chi_{\vec r \vec r'}^*\chi_{\vec r \vec r'} 
+ t_{\vec r \vec r'}^{\gamma *}t_{\vec r \vec r'}^\gamma 
+ e_{\vec r \vec r'}^{\gamma *}e_{\vec r \vec r'}^\gamma) \nonumber \\
&& +\frac{1}{8} \sum_{\gamma' \ne \gamma} (t_{\vec r \vec r'}^{\gamma' *}t_{\vec r \vec r'}^{\gamma'} 
+ e_{\vec r \vec r'}^{\gamma' *}e_{\vec r \vec r'}^{\gamma'} ), 
\label{eq:decoupling}
\end{eqnarray}
where 
$\Delta_{\vec r \vec r'} = f_{\vec r \sigma} i \tau^y_{\sigma \sigma'} f_{\vec r' \sigma'} $ (singlet pairing), 
$t^\gamma_{\vec r \vec r'} = f_{\vec r \sigma} [i \hat \tau^\gamma \hat \tau^y]_{\sigma \sigma'} f_{\vec r' \sigma'} $ 
(triplet pairing), 
$\chi_{\vec r \vec r'} = f_{\vec r \sigma}^\dag f_{\vec r' \sigma'}$ (spin-conserving exchange term), 
$e^\gamma_{\vec r \vec r'} = f_{\vec r \sigma}^\dag \tau^\gamma_{\sigma \sigma'} f_{\vec r' \sigma'}$ (spin-nonconserving exchange term). 
Summation over $\gamma$ in Eq.~(\ref{eq:decoupling}) gives a Heisenberg term. 
Then, the mean-field decoupling is introduced to terms having the negative coefficient. 
This recovers the previous mean-field schemes \cite{Lee06,Shindou09,Schaffer12}. 
Different decoupling schemes are also used in the literatures \cite{Burnell11,You11,Hyart12}. 
The full expression of the mean-field Hamiltonian is given in Ref.~\cite{supplement}.

We remark on the AFM Kitaev limit of the undoped $t_{2g}^5$ model. 
For this limit, we looked for self-consistent mean-field solutions which respect the underlying lattice symmetry. 
Such a solution was found to be given by 
$- \langle \chi_{x,y,z} \rangle = - \langle e^z_z\rangle = \langle t^x_x\rangle = i \langle t^y_y\rangle = 0.3812 i$ and 
$- \langle e^z_x\rangle = - \langle e^z_y\rangle = \langle t^x_y\rangle = \langle t^x_z\rangle 
= i \langle t^y_x\rangle = i \langle t^y_z\rangle = -0.1188 i$ with the other order parameters and 
the chemical potential being zero. 
Here, the notation is simplified by replacing the subscript $\vec r \vec r'$ with the bond index 
$\rho=x, y, z$ connecting the sites $\vec r \in A$ and $\vec r' \in B$; $\vec r_\rho = \vec r' - \vec r$. 
It is remarkable that this mean-field solution gives the spinon dispersion relation identical 
to that reported for the FM Kitaev model \cite{Schaffer12,supplement}; 
i.e., the ground state of the Kitaev model does not depend on the signs of the exchange constants \cite{Kitaev06}. 
The current {\it Ansatz} corresponds to the gauge used in Refs.~\cite{You11,Schaffer12}, 
and correctly describes a $Z_2$ SL.

{\it Doping effects}.---
We consider hopping matrices projected into neighboring Kramers doublets. 
In this representation, the hopping matrices are diagonal in the isospin index $\sigma$: 
$H_t = - \tilde t \sum_{\langle \vec r \vec r' \rangle \sigma} (c_{\vec r \sigma}^\dag c_{\vec r' \sigma} + H.c.)$. 
The hopping amplitude is renormalized according to the relative weight of the wave functions as 
$\tilde t = \frac{2}{3} (t_\pi + \frac{1}{2} t_\delta)$ [$\frac{1}{2} (t_\sigma + t_\delta)$] for the $t_{2g}^5$ [$e_g^{1,3}$] systems.
The double occupation is prohibited due to the strong repulsive interactions for $c$ operators. 
This effect at finite doping can be treated by introducing two bosonic auxiliary particles 
$b_{1,2}$ as 
$c_{\vec r \sigma} \Rightarrow \frac{1}{\sqrt{2}}
(b^\dag_{\vec r 1} f_{\vec r \sigma} + \sigma b^\dag_{\vec r 2} f_{\vec r \bar \sigma}^\dag )$ (Ref.~\cite{Lee06}) 
with the $SU(2)$ singlet condition $K_{\vec r}^\gamma = 
\frac{1}{4} {\rm Tr} \, 
F_{\vec r} \hat \tau^\gamma F^\dag_{\vec r} 
- \frac{1}{4} {\rm Tr} \, \hat \tau^z B^\dag_{\vec r} \hat \tau^\gamma B_{\vec r}= 0$. 
Here, 
$F_{\vec r} = 
\Bigl({f_{\vec r \uparrow} \atop f_{\vec r \downarrow}} 
{-f^\dag_{\vec r \downarrow} \atop f^\dag_{\vec r \uparrow}} \Bigr)$ 
and 
$B_{\vec r} = \Bigl({b^\dag_{\vec r 1} \atop b^\dag_{\vec r 2}} {-b_{\vec r 2} \atop b_{\vec r 1}} \Bigr)$ 
(Ref.~\cite{You11}), and 
the global constraints $\langle K^\gamma \rangle=0$ are imposed by 
$SU(2)$ gauge potentials $a^\gamma$. 
Doped carriers can be either holes or electrons, 
and the effect is symmetric for our model. 
We focus on the low-doping regime at zero temperature and
assume that all bosons are condensed, i.e., 
$\delta = \sum_\nu \langle b_{\nu \vec r}^\dag b_{\nu \vec r}\rangle 
\approx \sum_\nu |\langle b_{\nu \vec r} \rangle|^2 $ and 
$\langle b_{\nu \vec r \in A} \rangle = (\pm i) \langle b_{\nu \vec r' \in B} \rangle$, 
arriving at the mean-field hopping term: 
$H_t^{MF} = - \frac{\delta}{2} \tilde t \sum_{\langle \vec r \vec r' \rangle \sigma} 
\{(\mp i) f_{\vec r \sigma}^\dag f_{\vec r' \sigma} + H.c.\}$. 
The imaginary number $i$ arises when the Bose condensation has the sublattice-dependent phase \cite{You11}. 

Many mean-field parameters have to be solved self-consistently. 
In order to make the problem tractable, 
we focus on the following five {\it Ans{\"a}tze} which respect the sixfold rotational symmetry of the underlying lattice. 
The first {\it Ansatz}, termed $p$ SC$_1$, is adiabatically connected to the mean-field solution in the Kitaev limit given above. 
Here, the relative phase $\pm i$ is required between the Bose condensation at sublattices $A$ and $B$ 
with the $SU(2)$ gauge potentials $a^x=a^y=a^z$ \cite{You11,supplement}. 
The second {\it Ansatz} is a $p$ SC, termed $p$ SC$_2$, 
the third one is a singlet SC with the $s$ wave paring, 
and the fourth one is a singlet SC with the $d+id$ pairing.  
%
For the latter three {\it Ans{\"a}tze}, we further assume that  
(1) order parameters $\langle e^\gamma_\rho \rangle$ are zero because 
these indeed become zero at large dopings, 
(2) the bose condensation does not introduce a phase factor, and 
(3) the exchange term is symmetric $\langle \chi_\rho \rangle = \langle \chi \rangle$ and real. 
%
Thus, these {\it Ans{\"a}tze} are regarded as BCS-type weak coupling SCs. 
For the FM Ising case, 
magnetically ordered states with finite $\langle e^z_\rho \rangle = \langle e^z \rangle$ 
are considered as the fifth {\it Ansatz}. 
%

Because of the constraint $a^x=a^y=a^z$, 
the spinon density $\langle f_{\vec r \sigma}^\dag f_{\vec r \sigma} \rangle$ differs from 
the ``real'' electron density $\langle c_{\vec r \sigma}^\dag c_{\vec r \sigma}\rangle$ 
in the $p$ SC$_1$ phase and a normal phase 
($\langle t_\rho^\gamma \rangle = \langle e_\rho^\gamma \rangle = \langle \Delta_\rho \rangle =0$) 
adjacent to it. 
In many cases, such a normal phase has slightly lower energy than the other SC {\it Ans{\"a}tze}. 
We discard such a solution as it is an artifact by the constraint. 

\begin{figure}[tbp]
\includegraphics[width=0.7\columnwidth]{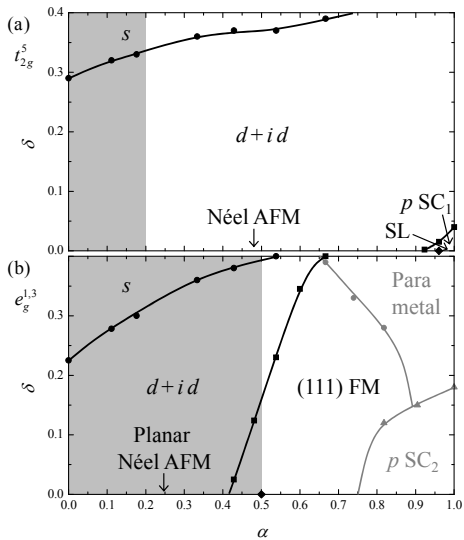}
\caption{Schematic phase diagrams for the doped AKAH model (a) and FIAH model (b) 
as a function of $\delta$ and $\alpha$. 
Parameters are taken as $J_K+J_H=t_\pi + \frac{1}{2} t_\delta$ and $J_I+J_H=t_\sigma + t_\delta$. 
Phase boundaries at finite $\delta$ are the results of the $SU(2)$ SBMF, 
while those at $\delta=0$ are results of the exact diagonalization. 
Shaded areas are the parameter ranges for $0<3I<U$ with $t_\delta=0$. 
Light lines in (b) are phase boundaries when the FM ordering is suppressed. 
}
\label{fig:x_alpha}
\end{figure}

The schematic phase diagrams for the doped AKAH model and FIAH model 
are shown in Figs.~\ref{fig:x_alpha} (a) and \ref{fig:x_alpha} (b), respectively, 
as a function of $\delta$ and $\alpha$. 
Here, to see various phases clearly, we chose the interaction strength as 
$J_K+J_H=t_\pi + \frac{1}{2} t_\delta$ and $J_I+J_H=t_\sigma + t_\delta$. 
For the AKAH model, 
the $p$ SC$_1$ phase is stabilized at $\alpha \sim 1$ and $\delta \sim 0$. 
Its area is quite small as its stability is intimately connected to that of the $Z_2$ spin liquid. 
The large area is covered by the singlet SCs, $d+id$ phases at small $\delta$ and $s$ at large $\delta$. 
This behavior results from the fact that the AFM Heisenberg term dominates the low-energy properties. 
%
For the FM Ising case, 
the (111) FM phase is stabilized in the large-$\alpha$ and small-$\delta$ regime. 
The $p$ SC$_2$ phase is also stabilized from the weak coupling mechanism 
but is found to exist only as a metastable phase. 

The $p$ SC$_1$ phase is characterized by the dispersive $\chi^0$ Majorana mode and the weakly dispersive $\chi^{x,y,z}$ modes. 
At finite $\delta$, 
all modes are gapped by the mixing between different Majorana modes due to the finite gauge potential $a^{x,y}$. 
This results in the finite Chern number +1. 

In the (111) FM, spin polarization is 100~\% at $\delta=0$ as in the exact diagonalization result. 
This large spin polarization persists up to relatively large $\delta$ as carriers can move without disturbing the spin ordering. 
The (111) FM area is extended to smaller $\alpha$ at $\delta \ne 0$ 
because the mean-field {\it Ansatz} for the (111) FM is closer to the true ground state at $\delta=0$ than that for the $d+id$. 
Since $\alpha$ is reduced by the direct $dd$ transfers, 
the unconventional $d+id$ SC is the most probable candidate induced by carrier doping.

{\it Discussion}.---
We now discuss the possible experimental realization of our model systems. 
A (111) bilayer of SrIrO$_3$ (Ref.~\cite{Cao07}) would be a good candidate for our AKAH model for $t_{2g}^5$ systems. 
Also, the FIAH model might be realized in a (111) bilayer of 
palladium oxide LaPdO$_3$ (Ref.~\cite{Kim01}). 
This $4d^7 (t_{2g}^6 e_g^1)$ electron system consists of nearly undistorted PdO$_6$ octahedra and 
is expected to have a stronger SOC than $3d$ counterparts such as LaNiO$_3$. 
Carrier doping would be achieved by partially substituting 
Ir by Ru or Os (hole doping) or Sr by La (electron doping) for SrIrO$_3$ and 
La by Sr (hole doping) or Pd by Ag or Au (electron doping) for LaPdO$_3$. 
It is yet to be clarified whether SrIrO$_3$ and LaPdO$_3$ are in the strong coupling regime, 
resulting in Mott insulators, 
or in the weak coupling regime, 
resulting in spin Hall insulators or topological metals \cite{Xiao11}. 
Even if these systems are in the Mott regime, the Kitaev SL may not be realized. 
But carrier doping would induce novel SC phases with $d+id$ symmetry.

For deriving effective models, the energy hierarchy is assumed as $U \gg \lambda \gg t$. 
Whether or not such a condition is realized in real materials remains to be examined. 
However, the effective transfer intensity is suppressed by correlations, and the corresponding hierarchy 
could be achieved self-consistently as discussed in Ref.~\cite{Pesin10}. 
(111) bilayers of perovskite oxides are plausible as the $d$ bands are relatively narrow 
(see, for example, band structures in Ref.~\cite{Xiao11}). 
The form of the nearest-neighbor interaction should not be altered even if the above hierarchy is broken as long as 
the local crystal field is maintained
and the interactions are expressed in terms of $S=1/2$ isospins 
because it relies on the symmetry and the spin conservation. 

Realizing $Z_2$ SL and $p$ SC phases may be preferable for fault tolerant topological computations. 
Within the current models, these phases are hard to achieve. 
For this purpose, an alternative route would be looking for systems with the FM Heisenberg interaction 
with which the parameter spaces for the $p$ SC phases in the doped systems 
are wider \cite{unpublished}.

To summarize, 
we studied the properties of Mott insulators realized in (111) bilayers of TMOs with a strong SOC. 
The low-energy effective models for such insulators consist of the anisotropic interaction and 
the AFM Heisenberg interaction.  
The former is of AFM Kitaev type for the $t_{2g}^5$ systems and FM Ising type for the $e_g^{1,3}$ systems. 
In both cases, large parameter spaces are characterized by magnetic long-range orderings 
with a narrow window for the SL regime in the $t_{2g}^5$ systems. 
Yet, carrier doping was found to make the physics of the current models more interesting 
by inducing unconventional SC phases in both cases. 
The most probable candidate is the singlet SC with the $d+id$ symmetry. 
In light of a weak SOC limit (Refs.~\cite{Ruegg11,Yang11}) 
and a strong coupling limit (Ref.~\cite{Kugel}), 
TMO (111) bilayers would provide even richer quantum behavior 
as a function of Coulomb interactions, the SOC and carrier doping.

We thank D. Xiao, Y. Ran, and G. Khaliullin for their fruitful discussions. 
This research was supported by 
the U.S. Department of Energy, Basic Energy Sciences, Materials Sciences and Engineering Division.


\newpage

\begin{center}
{\large \bf Supplementary material}





%
\end{center}
%

\renewcommand{\thetable}{S\Roman{table}}
\renewcommand{\thefigure}{S\arabic{figure}}
\renewcommand{\thesubsection}{S\arabic{subsection}}
\renewcommand{\thesubsubsection}{S\arabic{subsection}.\arabic{subsubsection}}
\renewcommand{\theequation}{S\arabic{equation}}

\setcounter{secnumdepth}{3}

\setcounter{equation}{0}
\setcounter{figure}{0}

\subsection{Model Hamiltonian}

\subsubsection{Transfer matrices}

For both $t_{2g}$ and $e_g$ systems, the hopping term is given by 
\begin{eqnarray}
H_t = -\sum_{\langle \vec r \vec r' \rangle} \sum_{o o' \sigma} 
\bigl( t_{\vec r \vec r'}^{o o'} d^\dag_{\vec r o \sigma} d_{\vec r' o' \sigma} + h. c. \bigr) 
\end{eqnarray}
where $d^\dag_{\vec r o \sigma}$ is the creation operator of an electron at site $\vec r$, orbital $o$ with spin $\sigma$. 
The hopping amplitude $t^{o o'}_{\vec r \vec r'}$ is determined from the Slater-Koster formula$^{21}$ 
with oxygen $p$ orbitals located between sites $\vec r$ and $\vec r'$. 

In $t_{2g}$ systems, electrons hop from site to site through $\pi$ bonding $(pd \pi)$ between 
the neighboring $t_{2g}$ orbitals and oxygen $2p$ orbitals in between 
and through weak direct overlap $(dd\delta)$. 
The dependence of the NN transfer matrices on the orbital and direction 
is given as follows: 
\begin{eqnarray}
t_{\vec r, \vec r \pm \hat{y} (\hat{z}) }^{a a} \!\!&=&\!\! 
t_{\vec r, \vec r \pm \hat{z} (\hat{x}) }^{b b} = 
t_{\vec r, \vec r \pm \hat{x} (\hat{y}) }^{c c} = t_\pi, \\
t_{\vec r, \vec r \pm \hat{x} }^{a a} \!\!&=&\!\! 
t_{\vec r, \vec r \pm \hat{y} }^{b b} = 
t_{\vec r, \vec r \pm \hat{z} }^{c c} = t_\delta, 
\end{eqnarray}
with the use of the following convention for the orbital index: 
$|a \rangle=|yz \rangle$, $|b \rangle=|zx \rangle$, and $|c \rangle=|xy \rangle$. 
$t_\pi \approx (pd \pi)^2/\Delta$ with $\Delta$ the level difference between TM $d$ orbitals and O $p$ orbitals, 
and $t_\delta =(dd\delta)$.

In $e_{g}$ systems, electrons hop from site to site through $\sigma$ bonding $(pd \sigma)$ between 
the neighboring $e_{g}$ orbitals and oxygen $2p$ orbitals and
and, similar to $t_{2g}$ systems, through weak direct overlap $(dd\delta)$. 
The dependence of the NN transfer matrices on the orbital and direction 
is given as follows: 
\begin{eqnarray}
t_{\vec r, \vec r \pm \hat{z}}^{\varepsilon \varepsilon'} \!\!&=&\!\! 
\left[
\begin{matrix}
t_{\vec r, \vec r \pm \hat{z}}^{\alpha \alpha} & t_{\vec r, \vec r \pm \hat{z}}^{\alpha \beta} \\
t_{\vec r, \vec r \pm \hat{z}}^{\beta \alpha} & t_{\vec r, \vec r \pm \hat{z}}^{\beta \beta}
\end{matrix}
\right]
=
\biggl[
\begin{matrix}
 t_\sigma & 0 \\
0 & t_\delta
\end{matrix}
\biggr], \\
t_{\vec r, \vec r \pm \hat{x}}^{\varepsilon \varepsilon'} \!\!&=&\!\! 
\frac{1}{4}
\left[
\begin{matrix}
t_\sigma + 3 t_\delta &  - \sqrt{3} (t_\sigma - t_\delta) \\
- \sqrt{3} (t_\sigma - t_\delta) & 3 t_\sigma + t_\delta
\end{matrix}
\right], \\
t_{\vec r, \vec r \pm \hat{y}}^{\varepsilon \varepsilon'} \!\!&=&\!\! 
\frac{1}{4}
t_\sigma \left[
\begin{matrix}
t_\sigma + 3 t_\delta &  \sqrt{3} (t_\sigma - t_\delta) \\
\sqrt{3} (t_\sigma - t_\delta) & 3 t_\sigma + t_\delta
\end{matrix}
\right], 
\end{eqnarray}
with the basis 
$| \alpha \rangle = | 3z^2-r^2 \rangle$ and $| \beta \rangle = | x^2-y^2 \rangle$. 
$t_\sigma \approx (pd \sigma)^2/\Delta$.

\subsubsection{Spin-orbit coupling}

The SOC for the $t_{2g}$ model is given by 
\begin{equation}
H^{t_{2g}}_{SO} =
\frac{\lambda}{2} \sum_{\vec r \sigma \sigma'} 
\sum_{{\scriptstyle \tau \tau' \tau''}\atop{\scriptstyle \in t_{2g}}}
i \varepsilon_{\tau \tau' \tau''} d^\dag_{\vec r \tau \sigma} 
\sigma^{\tau''}_{\sigma \sigma'} 
d_{\vec r \tau' \sigma}, 
\label{eq:Hso}
\end{equation}
where 
$\varepsilon_{\tau \tau' \tau''}$ is the Levi-Civita antisymmetric tensor.

In (111) bilayers, the SOC in the $e_g$ multiplet is activated through the virtual electron excitation to 
the $t_{2g}$ multiplet under the trigonal $C_{3v}$ crystalline field.$^{16}$ 
The resulting SOC is expressed as 
\begin{eqnarray}
H^{e_g}_{SO} = - \tilde \lambda \sum_{\vec r \sigma} \sum_{\varepsilon \varepsilon' \in e_g} 
d^\dag_{\vec r \varepsilon \sigma}
\tau^y_{\varepsilon \varepsilon'} \tau^z_{\sigma \sigma} d_{\vec r \varepsilon' \sigma}, 
\label{eq:HegSO}
\end{eqnarray}
where $\hat \tau$ are Pauli matrices. 
Here, the spin quantization axis is taken along the [111] crystallographic axis. 
By diagonalizing the Hamiltonian Eq.~(\ref{eq:HegSO}), one obtains the Kramers doublet given by 
Eq.~(3). 

\subsubsection{Local Coulomb interactions}

For simplicity, we neglect the coupling between $t_{2g}$ electrons and $e_g$ electrons 
in the local interaction. 
Thus, the multiorbital interaction for both the cases can be expressed as 
\begin{equation}
H_U =
\frac{1}{2} \sum_{\vec r}
\sum_{{\scriptstyle  o o'}\atop{\scriptstyle o'' o'''}} \sum_{\sigma \sigma'} 
U^{o o' o'' o'''} 
d_{\vec r o \sigma}^\dag d_{\vec r o' \sigma'}^\dag d_{\vec r o''' \sigma'} d_{\vec r o'' \sigma}, 
\label{eq:HU}
\end{equation}
where the orbital indices $o, \ldots o'''$ run through either the $t_{2g}$ multiplet or the $e_g$ multiplet 
(see for example Ref.~S1). 
Because of the orbital symmetry, 
a well know relation $U=U'+2I$ holds, where 
$U=U^{o o o o}$ (intraorbital Coulomb), 
$U'=U^{o o' o o'}$ (interorbital Coulomb), 
$I=U^{o o' o' o}$ (interorbital exchange) 
$= U^{o o o' o'}$ (interorbital pair transfer) for $o \ne o'$, 
and other components are absent. 

Equation~(\ref{eq:HU}) can be easily diagonalized when two electrons occupy site $r$.  
Resulting energy eigenstates and eigenvalues are as follows: 
\begin{eqnarray}
\left\{ 
\begin{array}{rcl}
^3T_2: & U-3I & (\mbox{nine-fold degenerate}) \\
^3T_1: & U-I  &(\mbox{three-fold degenerate}) \\
^1T_2: & U-I  &(\mbox{two-fold degenerate}) \\
^1T_1: & U+2I & (\mbox{non degenerate}) \\
\end{array}
\right.
\end{eqnarray}
for $t_{2g}^2$, 
and
\begin{eqnarray}
\left\{ 
\begin{array}{rcl}
^3A_2: & U-3I & (\mbox{three-fold degenerate}) \\
^1E: & U-I  &(\mbox{two-fold degenerate}) \\
^1A_2: & U+I & (\mbox{non degenerate}) \\
\end{array}
\right.
\end{eqnarray}
for $e_{g}^2$. 
These energy levels determine the excitation energy for $t_{2g}^1 t_{2g}^1 \rightleftharpoons t_{2g}^2 t_{2g}^0$ 
and $e_{g}^1 e_{g}^1 \rightleftharpoons e_{g}^2 e_{g}^0$, and also 
$t_{2g}^5 t_{2g}^5 \rightleftharpoons t_{2g}^6 t_{2g}^4$ 
and $e_{g}^3 e_{g}^3 \rightleftharpoons e_{g}^4 e_{g}^2$ by the particle-hole symmetry. 

\subsubsection{Effective interactions}

Considering the limit $U \gg \lambda (\widetilde \lambda) \gg t_{\pi (\sigma)}$ and 
$U \gg (\mbox{trigonal crystal field splitting for $t_{2g}$ multiplet})$, 
we include the SOC in the initial states and the final states of the second-order perturbation with respect to $H_t$, 
arriving at the effective Hamiltonian Eq.~(2) 
for $t_{2g}^5$ systems and 
Eq.~(4) 
for $e_g^{1,3}$ systems. 

Here we provide full expressions for the effective interactions for $t_{2g}^5$ systems [Eq.~(2)] and $e_g^{1,3}$ systems [Eq.~(4)]. 
For $t_{2g}^5$ systems, we obtain 
\begin{eqnarray}
J_K \!\!&=&\!\! \frac{4}{9} (1-2\nu+\nu^2) (r_1 -  r_2),\\
J_H \!\!&=&\!\! \frac{8}{27} \{ ( 3 + 6 \nu ) r_1 
+(1-2 \nu + \nu^2 ) r_2 
+( 2 + 2 \nu + \nu^2 ) r_3 \} , \nonumber \\
\end{eqnarray}
with $r_1 = t_\pi^2/(U-3I)$, $r_2 = t_\pi^2/(U-I)$, $r_3 = t_\pi^2/(U+2I)$, and $\nu = t_\delta/t_\pi$. 
For $e_g^{1,3}$ systems, we obtain 
\begin{eqnarray}
J_I \!\!&=&\!\! \frac{1}{2} \{ (2-4\nu + 2\nu^2 ) r_1 - (1 - 4\nu +\nu^2) r_2 - (1+\nu^2) r_3 \}, \nonumber \\ \\
J_H \!\!&=&\!\! \frac{1}{2} (1+2\nu+\nu^2)(r_1 + r_3),
\end{eqnarray}
with 
$r_1 = t_\sigma^2/(U-3I)$, $r_2 = t_\sigma^2/(U-I)$, $r_3 = t_\sigma^2/(U+I)$, 
and $\nu = t_\delta/t_\sigma$.

We check these interactions by considering two limiting cases. 
(i) $I \rightarrow 0$, both $J_K$ and $J_I$ become zero. 
This is because, in the intermediate states of the second-order perturbation processes, 
interorbital contributions, 
sum of $^3T_2$ and $^3T_1$ for $t_{2g}^5$ and sum of $^3A_2$ and $^1E$ for $e_g^{1,3}$, 
vanish and only intraorbital contributions remain. 
Intraorbital contributions involve configurations such as $|a_\uparrow a_\downarrow \rangle$, 
resulting in the AF interactions $J_H$. 
(ii) $t_\delta \rightarrow t_\pi$ or $t_\delta \rightarrow t_\sigma$, 
the directionality coming from $d$-orbital wave functions is lost. 
Thus, in this case, $J_K$ vanishes for the $t_{2g}^5$ model. 
On the other hand, $J_I$ remains finite for the $e_{g}^{1,3}$ model. 
This is because the total $\vec S \parallel [111]$ is conserved.

\subsection{Mean field Hamiltonians}

After the mean-field decoupling, the single-particle Hamiltonian for 
the AF Kitaev-AF Heisenberg model for $t_{2g}^{5}$ systems 
is expressed as 
\begin{eqnarray}
H^{MF}=\sum_{\vec k} \sum_{\sigma \sigma'} \varphi_{\vec k \sigma}^\dag \hat H (\vec k) \varphi_{\vec k \sigma'}
+H_0.
\label{eq:HMF}
\end{eqnarray}
Here, a Nambu representation is used with 
4-component spinors $\varphi^\dag_{\vec k \sigma}$ given by 
$\varphi^\dag_{\vec k \sigma} = 
\bigl(f^\dag_{\vec k A \sigma}, f^\dag_{\vec k B \sigma}, f_{-\vec k A \sigma}, f_{-\vec k B \sigma}\bigr)$, 
and an $8 \times 8$ matrix $\hat H$ given by 
\begin{widetext}
\begin{eqnarray}
\hat H (\vec k) = 
\left[
\begin{array}{cccc}
- a^z \, \delta_{\sigma \sigma'} & \chi_{\sigma \sigma'} (\vec k) & (a^x + i a^y) \varepsilon_{\sigma \sigma'} 
& \Delta_{\sigma \sigma'} (\vec k) \\
\chi^*_{\sigma' \sigma} (\vec k) & - a^z \, \delta_{\sigma' \sigma} & -\Delta_{\sigma' \sigma} (-\vec k) & 
(a^x + i a^y) \varepsilon_{\sigma \sigma'} \\
(a^x - i a^y) \varepsilon_{\sigma' \sigma}& -\Delta^*_{\sigma \sigma'} (-\vec k) & a^z \, \delta_{\sigma \sigma'} & 
-\chi^*_{\sigma' \sigma} (-\vec k)\\
\Delta^*_{\sigma' \sigma} (\vec k) & (a^x - i a^y) \varepsilon_{\sigma' \sigma} & -\chi^*_{\sigma \sigma'} (-\vec k) & 
a^z \, \delta_{\sigma \sigma'}
\end{array}
\right]. 
\end{eqnarray}
$\varepsilon_{\uparrow \downarrow}=-\varepsilon_{\downarrow \uparrow} = 1$ is the antisymmetric tensor. 
$\hat \chi (\vec k)$ and $\hat \Delta (\vec k)$ are $2 \times 2$ matrices given by 
\begin{eqnarray}
\hat \chi (\vec k) \!\!&=&\!\! - \frac{1}{8} \sum_\rho e^{i \vec k \cdot \vec r_\rho} 
\bigl\{ 4 \delta (i) \tilde t + (J_K + 3 J_H)\langle \chi_\rho^* \rangle \bigr\} \hat \tau^0 
- \frac{1}{8} \sum_{\rho} e^{i \vec k \cdot \vec r_\rho} 
J_K \langle e^{\rho *}_\rho  \rangle 
\,
\hat \tau^\rho ,\\
\hat \Delta (\vec k) \!\!&=&\!\! \frac{1}{8} \sum_{\rho} e^{i \vec k \cdot \vec r_\rho}
(J_K + 3 J_H) \langle \Delta_\rho \rangle i \hat \tau^y 
-\frac{1}{8} \sum_{\rho} e^{i \vec k \cdot \vec r_\rho} J_K 
\langle t^{\rho}_\rho \rangle \,
i \hat \tau^y \hat \tau^\rho, 
\label{eq:Delta}
\end{eqnarray}
respectively, 
with $\hat \tau^0$ being the $2 \times 2$ unit matrix. 
$\vec r_\rho$ is a unit vector connecting the nearest-neighboring sites along the $\rho$ bond, i.e.,
$\vec r_x = (-\sqrt{3}/2,-1/2)$, $\vec r_y = (\sqrt{3}/2,-1/2)$ and 
$\vec r_z = (0,1)$. 
The prefactor for $\tilde t$ comes from the mean-field decoupling for the bosonic term 
$\langle b_{A 1} b^\dag_{B 1} -  b^\dag_{A 2} b_{B 2}  \rangle$. 
For the $p$ SC$_1$ phase, the bose condensation at one of the two sublattices acquires phase $i$, 
thus we have $\langle b_{A 1} b^\dag_{B 1} -  b^\dag_{A 2} b_{B 2} \rangle \approx 
\pm i \{ \langle b_{1} \rangle^2 + \langle b_{2} \rangle^2\} = \pm i \delta $, 
while for the other phases considered, only $b_1$ bosons condense at finite $\delta$, thus 
$\langle b_{A 1} b^\dag_{B 1} -  b^\dag_{A 2} b_{B 2}  \rangle \approx \langle b_{1} \rangle^2 = \delta$. 
$H_0$ is a constant term given by 
$H_0 = \frac{1}{8}\sum_{\rho} J_K \bigl(|\langle \chi_\rho \rangle|^2 + |\langle e^\rho_\rho \rangle|^2 
+ |\langle \Delta_\rho \rangle|^2 + |\langle t^\rho_\rho \rangle|^2 \bigr) 
+ \frac{3}{8} \sum_{\rho} J_H \bigl(|\langle \chi_\rho \rangle|^2 + |\langle \Delta_\rho \rangle|^2 \bigr)$.

For the FM Ising-AF Heisenberg model for $e_{g}^{1,3}$ systems, 
including the uniform magnetic moment $m = \langle f^\dag_{i \uparrow} f_{i \uparrow} - f^\dag_{i \downarrow} f_{i \downarrow} \rangle$, 
we have 
\begin{eqnarray}
\hat H (\vec k) = 
\left[
\begin{array}{cccc}
- a^z \, \delta_{\sigma \sigma'} - \frac{3}{4} J_I m \sigma_{\sigma \sigma'}^z & \chi_{\sigma \sigma'} (\vec k) & (a^x + i a^y) \varepsilon_{\sigma \sigma'} 
& \Delta_{\sigma \sigma'} (\vec k) \\
\chi^*_{\sigma' \sigma} (\vec k) & - a^z \, \delta_{\sigma \sigma'} - \frac{3}{4} J_I m \sigma_{\sigma \sigma'}^z & -\Delta_{\sigma' \sigma} (-\vec k) & 
(a^x + i a^y) \varepsilon_{\sigma \sigma'} \\
(a^x - i a^y) \varepsilon_{\sigma' \sigma}& -\Delta^*_{\sigma \sigma'} (-\vec k) & a^z \, \delta_{\sigma \sigma'} + \frac{3}{4} J_I m \sigma_{\sigma \sigma'}^z & 
-\chi^*_{\sigma' \sigma} (-\vec k)\\
\Delta^*_{\sigma' \sigma} (\vec k) & (a^x - i a^y) \varepsilon_{\sigma' \sigma} & -\chi^*_{\sigma \sigma'} (-\vec k) & 
a^z \, \delta_{\sigma \sigma'} + \frac{3}{4} J_I m \sigma_{\sigma \sigma'}^z
\end{array}
\right] 
\end{eqnarray}
with
%
%
\begin{eqnarray}
\hat \chi (\vec k) \!\!&=&\!\! - \frac{1}{8} \sum_\rho e^{i \vec k \cdot \vec r_\rho} 
[ \bigl\{ 4 \delta (i) \tilde t + 3 J_H \langle \chi_\rho^* \rangle \bigr\} \hat \tau^0 - J_H \langle e_\rho^{z *} \rangle \hat \tau^z]
- \frac{1}{8} \sum_{\rho} \sum_{\gamma=x,y} e^{i \vec k \cdot \vec r_\rho} 
J_I \langle e^{\gamma *}_\rho \rangle 
\,
\hat \tau^\gamma ,\\
\hat \Delta (\vec k) \!\!&=&\!\! \frac{3}{8} \sum_{\rho} e^{i \vec k \cdot \vec r_\rho}
J_H \langle \Delta_\rho \rangle i \hat \tau^y 
-\frac{1}{8} \sum_{\rho} \sum_{\gamma = x,y} e^{i \vec k \cdot \vec r_\rho} 
J_I \langle t^{\gamma}_\rho \rangle
\,
i \hat \tau^y \hat \tau^\gamma, 
\end{eqnarray}
and the constant term $H_0$ is given by 
$H_0 = \frac{1}{8}\sum_{\rho} \sum_{\gamma=x,y} J_I \bigl(|\langle e^\gamma_\rho \rangle|^2 
+ |\langle t^\gamma_\rho \rangle |^2 \bigr) 
+ \frac{1}{8} \sum_{\rho} J_H \bigl(3 |\langle \chi_\rho \rangle|^2 -  |\langle e_\rho^z \rangle|^2 + 3 |\langle \Delta_\rho \rangle|^2 \bigr)$. 
%
\end{widetext}

\subsection{Majorana representation for the $p$ SC$_1$ phase in the AF Kitaev-AF Heisenberg model}

For the $p$ SC$_1$ phase, a mean field solution which respect the underlying lattice symmetry is
given by 
$\langle \chi_{x,y,z} \rangle = -i A'$, 
$-\langle e^z_z\rangle = \langle t^x_x\rangle = i\langle t^y_y\rangle = i A$ and 
$-\langle e^z_x\rangle = -\langle e^z_y\rangle = \langle t^x_y\rangle = \langle t^x_z\rangle 
= i \langle t^y_x\rangle = i \langle t^y_z\rangle = i B$, 
where $A,A',B$ are real numbers. 
Spinon operators can be represented as linear combinations of Majorana fermions. 
Here, we use the gauge given by 
You {\it et al.}:$^{11}$ 
\begin{eqnarray}
f_\uparrow \!\!&=&\!\! \frac{1}{\sqrt{2}}(\chi^0 + i \chi^z), \\ 
f_\uparrow^\dag \!\!&=&\!\! \frac{1}{\sqrt{2}}(\chi^0 - i \chi^z), \\
f_\downarrow \!\!&=&\!\! \frac{1}{\sqrt{2}}(i \chi^x - \chi^y), \\
f_\downarrow^\dag \!\!&=&\!\! \frac{1}{\sqrt{2}}(-i \chi^x - \chi^y). 
\end{eqnarray}
Spin operators are represented by these Majorana fermions as $S^\gamma = i \chi^0 \chi^\gamma$,
with the local constraint $\chi^0 \chi^x \chi^y \chi^z = 1/4$ and 
the normalization $\{ \chi^\gamma, \chi^{\gamma'} \} = \delta_{\gamma \gamma'}$.

\begin{widetext}
By inserting the mean field parameters and spinon operators in terms of Majorana fermion operators, 
we obtain the following expressions for the mean-field Hamiltonian along the $x, y$, and $z$ directions: 
\begin{eqnarray}
H^{MF}_x \!\!&=&\!\!-\frac{1}{8} \bigl\{ 4 \tilde t i \delta +(J_K + 3J_H) i A' \bigr\}(\chi^0_A \chi^0_B + \chi^x_A \chi^x_B 
+ \chi^y_A \chi^y_B + \chi^z_A \chi^z_B) 
-\frac{1}{8} J_K i A (\chi^0_A \chi^0_B + \chi^x_A \chi^x_B 
- \chi^y_A \chi^y_B - \chi^z_A \chi^z_B), \nonumber \\ \\
H^{MF}_y \!\!&=&\!\!-\frac{1}{8} \bigl\{ 4 \tilde t i \delta + (J_K + 3 J_H) i A' \bigr\}(\chi^0_A \chi^0_B + \chi^x_A \chi^x_B 
+ \chi^y_A \chi^y_B + \chi^z_A \chi^z_B) 
-\frac{1}{8} J_K i A (\chi^0_A \chi^0_B - \chi^x_A \chi^x_B 
+ \chi^y_A \chi^y_B - \chi^z_A \chi^z_B), \nonumber \\ \\
H^{MF}_z \!\!&=&\!\!-\frac{1}{8} \bigl\{ 4 \tilde t i \delta + (J_K +3 J_H) i A' \bigr\} (\chi^0_A \chi^0_B + \chi^x_A \chi^x_B 
+ \chi^y_A \chi^y_B + \chi^z_A \chi^z_B) 
-\frac{1}{8} J_K i A (\chi^0_A \chi^0_B - \chi^x_A \chi^x_B 
- \chi^y_A \chi^y_B + \chi^z_A \chi^z_B). \nonumber \\
%
\end{eqnarray}
\end{widetext}

In the undoped Kitaev limit ($\delta=0$ and $J_H=0$), $A=A'$ and, therefore, 
this mean field Hamiltonian is reduced to 
\begin{eqnarray}
H^{MF}_\rho = 
-\frac{1}{4} J_K i A(\chi^0_A \chi^0_B + \chi^\rho_A \chi^\rho_B). 
\end{eqnarray}
Thus, we recover the correct Kitaev limit 
with the dispersive $\chi^0$ mode and the dispersionless $\chi^{x,y,z}$ modes. 
With finite $J_H$, $A' \ne A$, and 
the $\chi^{x,y,z}$ modes acquire finite dispersions.

At finite $\delta$, 
the gauge potentials have to be explicitly included as $a^z$ acts as the chemical potential for spinons. 
The local term involving the gauge potentials is given in terms of spinons or Majorana fermions as 
\begin{eqnarray}
&& \hspace{-2em}
\sum_\gamma a^\gamma \frac {1}{4} {\rm Tr} \, \biggl(
F_{\vec r} \hat \tau^\gamma F^\dag_{\vec r} 
- \hat \tau^z B^\dag_{\vec r} \hat \tau^\gamma B_{\vec r} \biggr) 
\nonumber \\
&&= \frac{1}{2} a^x \bigl( -i \chi^0 \chi^x - i \chi^y \chi^z - b^\dag_1 b_2 -b^\dag_2 b_1 \bigr) \nonumber \\
&& \quad + \frac{1}{2} a^y \bigl( -i \chi^0 \chi^y - i \chi^z \chi^x + i b^\dag_1 b_2 - i b^\dag_2 b_1 \bigr) \nonumber \\
&& \quad + \frac{1}{2} a^z \bigl( -i \chi^0 \chi^z - i \chi^x \chi^y - b^\dag_1 b_1 + b^\dag_2 b_2 \bigr). 
\end{eqnarray}
Thus, in order to have the correct lattice symmetry ($x \leftrightarrow y \leftrightarrow z$) at finite doping, 
all gauge potentials must have the equal absolute value. 
At the same time, the mixing between different Majorana modes generates excitation gaps, 
resulting in the finite Chern number.$^{11}$ 

\begin{figure}[htbp]
\includegraphics[width=0.8\columnwidth]{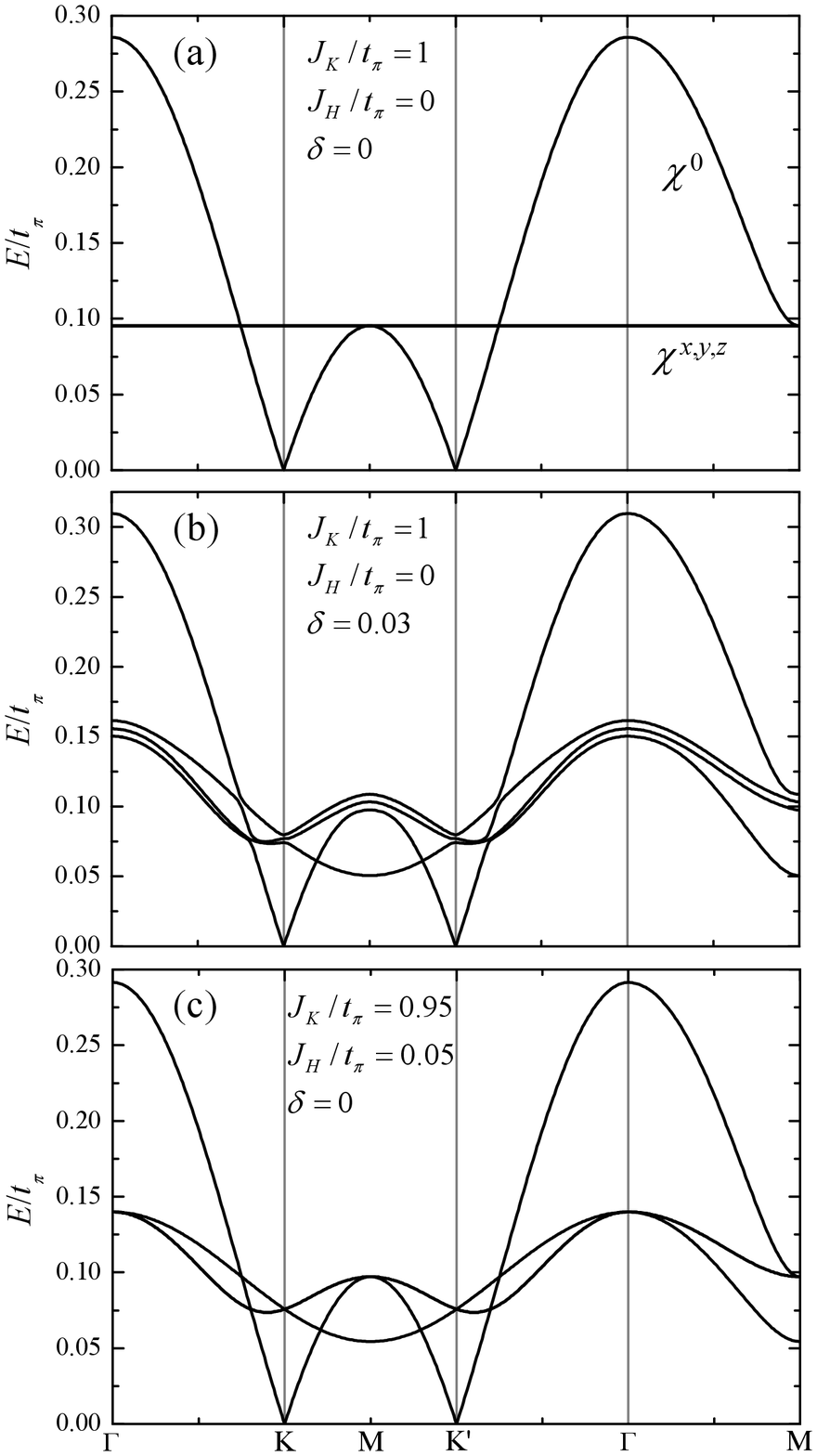}
\caption{Dispersion relations of the Majorana fermions for the AF Kitaev-AF Heisenberg model. 
(a) Undoped Kitaev limit, (b) doped Kitaev, (c) undoped Kitaev-Heisenberg. 
Parameter values are indicated.}
\label{fig:dispersion}
\end{figure}

Dispersion relations of the Majorana fermions are presented in Fig.~\ref{fig:dispersion} for various choices of parameters. 
In the undoped Kitaev model (a), only the gapless $\chi^0$ mode is dispersive. 
With finite doping $\delta$ (b), $\chi^{x,y,z}$ modes become dispersive and the $\chi^0$ mode is gapped. 
With finite $J_H$ (c), $\chi^{x,y,z}$ modes become dispersive while the $\chi^0$ mode remains gapless. 
With $J_K=t_\pi, J_H=0$ and $\delta=0.03$, the gap amplitude is $\sim 3 \times 10^{-5} t_\pi$ and, therefore, 
invisible in this scale. 
Softening of the $\chi^{x,y,z}$ modes is not strong enough to close a gap. 
This results in the Chern number $+1$.

\vspace{1em}

\textbf{References}
\begin{itemize}
\item[S1]
S. Sugano, Y. Tanabe, and H. Kamimura, {\it Multiplets of Transition-Metal Ions in Crystals} 
(Academic, New York, 1970). 

\end{itemize}


\begin{thebibliography}{*}

\bibitem{Kim08}B. J. Kim, H. Jin, S. J. Moon, J.-Y. Kim, B.-G. Park, C. S. Leem, J. Yu, T. W. Noh, C. Kim, S.-J. Oh,
J.-H. Park, V. Durairaj, G. Cao, and E. Rotenberg, Phys. Rev. Lett. {\bf 101}, 076402 (2008). 

\bibitem{Pesin10}D. Pesin and L. Balents, Nat. Phys. {\bf 6}, 376 (2010).

\bibitem{Wang11}F. Wang and T. Senthil, Phys. Rev. Lett. {\bf 106}, 136402 (2011). 

\bibitem{Shitade09}A. Shitade, H. Katsura, J. Kune{\v s}, X.-L. Qi, S.-C. Zhang, and N. Nagaosa, 
Phys. Rev. Lett. {\bf 102}, 256403 (2009). 

\bibitem{Chaloupka10}J. Chaloupka, G. Jackeli, and G. Khaliullin, Phys. Rev. Lett. {\bf 105}, 027204 (2010). 

\bibitem{Kitaev06}A. Kitaev, Ann. Phys. (N.Y.) {\bf 321}, 2 (2006).

\bibitem{Singh10}Y. Singh and P. Gegenwart, Phys. Rev. B {\bf 82}, 064412 (2010). 

\bibitem{Liu11}X. Liu, T. Berlijn, W.-G. Yin, W. Ku, A. Tsvelik, Y.-J. Kim, H. Gretarsson, 
Y. Singh, P. Gegenwart, and J. P. Hill, Phys. Rev. B {\bf 83}, 220403 (2011). 

\bibitem{Ye12}
F. Ye, S. Chi, H. Cao, B. C. Chakoumakos, J. A. Fernandez-Baca, R. Custelcean, T. F. Qi, O.B. Korneta, and G. Cao, 
Phys. Rev. B {\bf 85}, 180403 (2012). 

\bibitem{Singh12}
Y. Singh, S. Manni, J. Reuther, T. Berlijn, R. Thomale, W. Ku, S. Trebst, and P. Gegenwart, 
Phys. Rev. Lett. {\bf 108}, 127203 (2012). 

\bibitem{Choi12}
S. K. Choi, R. Coldea, A. N. Kolmogorov, T. Lancaster, I. I. Mazin, S. J. Blundell, P. G. Radaelli, Y. Singh, 
P. Gegenwart, K. R. Choi, S.-W. Cheong, P. J. Baker, C. Stock, and J. Taylor, 
Phys. Rev. Lett. {\bf 108}, 127204 (2012).  



\bibitem{Kimchi11}I. Kimchi and Y.-Z. You, Phys. Rev. B {\bf 84}, 180407 (2011) 



\bibitem{You11}Y.-Z. You, I. Kimchi, and A. Vishwanath, Phys. Rev. B, {\bf 86}, 085145 (2012). 
\bibitem{Hyart12}T. Hyart, A. R. Wright, G. Khaliullin, and B. Rosenow, Phys. Rev. B {\bf 85}, 140510 (2012). 

\bibitem{Haldane88}F. D. M. Haldane, Phys. Rev. Lett. {\bf 61}, 2015 (1988).

\bibitem{Kane05}C. L. Kane and E. J. Mele  Phys. Rev. Lett. {\bf 95}, 146802 (2005).


\bibitem{Raghu08}S. Raghu, X.-L. Qi, C. Honerkamp, and S.-C. Zhang, Phys. Rev. Lett. {\bf 100}, 156401 (2008). 


\bibitem{Xiao11}D. Xiao, W. Zhu, Y. Ran, N. Nagaosa, and S. Okamoto, Nat. Commun. {\bf 2}, 596 (2011). 

\bibitem{Ruegg11}A. R{\"u}egg and G. A. Fiete, Phys. Rev. B {\bf 84}, 201103 (2011). 
\bibitem{Yang11}K.-Y. Yang, W. Zhu, D. Xiao, S. Okamoto, Z. Wang, and Y. Ran, Phys. Rev. B {\bf 84}, 201104 (2011). 

\bibitem{Hwang12}For the latest development, see H. Y. Hwang, Y. Iwasa, M. Kawasaki, B. Keimer, N. Nagaosa, and Y. Tokura
Nat. Mater. {\bf 11}, 103(2012). 

\bibitem{t2g1}With the strong SOC, $t_{2g}^1$ systems are described in terms of effective angular momentum $J_{eff}=3/2$. 

\bibitem{Slater54}J. C. Slater and G. F. Koster, Phys. Rev. {\bf 94}, 1498 (1954).

\bibitem{supplement}See Supplemental Material for more information.

\bibitem{Jackeli09}G. Jackeli and G. Khaliullin, Phys. Rev. Lett. {\bf 102}, 017205 (2009).

\bibitem{Khaliullin05}G. Khaliullin, Prog. Theor. Phys. Suppl. {\bf 160}, 155 (2005).

\bibitem{Chaloupka12}J. Chaloupka, G. Jackeli, and G. Khaliullin, arXiv:1209.5100.

\bibitem{unpublished}S. Okamoto, arXiv:1212.5218. 

\bibitem{Black07}A. M. Black-Schaffer and S. Doniach, Phys. Rev. B {\bf 75}, 134512 (2007). 


\bibitem{Lee06}P. A. Lee, N. Nagaosa, and X.-G. Wen, Rev. Mod. Phys. {\bf 78}, 17 (2006). 

\bibitem{Shindou09}R. Shindou and T. Momoi, Phys. Rev. B {\bf 80}, 064410 (2009). 

\bibitem{Schaffer12}R. Schaffer, S. Bhattacharjee, and Y.-B. Kim, Phys. Rev. B {\bf 86}, 224417 (2012).

\bibitem{Burnell11}F. J. Burnell and C. Nayak, Phys. Rev. B {\bf 84}, 125125 (2011). 




\bibitem{Cao07}G. Cao,  V. Durairaj, S. Chikara, L. E. DeLong, S. Parkin, and P. Schlottmann, 
Phys. Rev. B {\bf 76}, 100402(R) (2007).

\bibitem{Kim01}S.-J. Kim , S. Lemaux, G. Demazeau, J.-Y. Kim, and J.-H. Choy, J. Am. Chem. Soc. {\bf 123}, 10413 (2001); 
J. Mater. Chem. {\bf 12}, 995 (2002). 


\bibitem{Kugel}K. I. Kugel and D. I. Khomskii, Sov. Phys. Usp. {\bf 25}, 231 (1982); Sov. Phys. Solid State {\bf 17}, 285 (1975).


\end{thebibliography}
\end{document}